\documentclass[preprint]{aastex}

\slugcomment{To appear in {\sl The Astrophysical Journal}}

\begin{document}

\title{An Accreting Black Hole in the Nuclear Star Cluster of the Bulgeless 
Galaxy NGC~1042}

\author{Joseph C. Shields\altaffilmark{1}, C. Jakob Walcher\altaffilmark{2,3}, 
Torsten B{\"o}ker\altaffilmark{4}, Luis C. Ho\altaffilmark{5}, 
Hans-Walter Rix\altaffilmark{6}, and Roeland P. van der Marel\altaffilmark{7}}

\altaffiltext{1}{Ohio University, Physics and Astronomy Department, 
Clippinger Labs, Athens OH 45701-2979, USA}
\altaffiltext{2}{Laboratoire d'Astrophysique de Marseille, 
Traverse du Siphon-Les trois Lucs, 13376 Marseille Cedex 12, France}
\altaffiltext{3}{Institut d'Astrophysique de Paris, 98bis, Bd. Arago,
75014 Paris, France}
\altaffiltext{4}{ESA/ESTEC, Keplerlaan 1, 2200 AG Noordwijk, Netherlands}
\altaffiltext{5}{The Observatories of the Carnegie Institution of Washington,
813 Santa Barbara Street, Pasadena, CA 91101-1291, USA}
\altaffiltext{6}{Max-Planck-Institut f{\"u}r Astronomie, K{\"o}nigstuhl 17,
D-69117 Heidelberg, Germany}
\altaffiltext{7}{Space Telescope Science Institute, 3700 San Martin Dr., 
Baltimore, MD 21218, USA}

\begin{abstract}
We present spectroscopic evidence for a low-luminosity, low excitation active
galactic nucleus (AGN) in NGC~1042, powered by an intermediate-mass
black hole.  These findings are significant in that the AGN is
coincident with a compact star cluster known to reside in the
nucleus, thus providing an example where the two types of central
mass concentration coexist.  The existence of a central black hole
is additionally remarkable in that NGC~1042 lacks a stellar bulge.
Objects such as NGC~1042 may have an important role in testing
theories for the genesis of massive black holes in galaxy nuclei,
and the extent to which they are in symbiosis with the larger
stellar host.
\end{abstract}

\keywords{galaxies: nuclei --- galaxies: active --- galaxies: star clusters}

\section{Introduction}

The origins of supermassive black holes in galaxy nuclei are not well
understood.  While core-collapse supernovae provide a natural means of
producing stellar-mass black holes, dissipation on a much larger scale
is required to build up black holes with masses of $M_\bullet \approx
10^7 - 10^{9}$ M$_\odot$ as inferred in luminous quasars and in the
centers of massive quiescent galaxies.  The gas accretion that powers
active galactic nuclei is a means of growing black hole mass, but the
onset of QSO activity at redshifts $z > 6$ and the correspondingly
short timescale for black hole growth may implicate another growth
mechanism that is rapid and radiatively inefficient.  These
considerations have prompted interest in mechanisms for producing
``seed'' black holes of intermediate mass ($\sim 10^2 - 10^6$ M$_\odot$)
that later emerge as the drivers of luminous AGNs 
(see Shapiro 2004 and van der Marel 2004 for reviews).

Dense star clusters provide one possible vehicle for generating
intermediate mass black holes \citep[e.g.,][and references therein]
{rasio04}.  Interest in star clusters as possible precursers for
massive black holes has grown as a result of surveys demonstrating
that clusters are commonly found in the centers of disk galaxies
\citep{phillips96, carollo98, boeker02, seth06} and also ellipticals
\citep{grant05, cote06}.  A physical connection between nuclear star
clusters and supermassive black holes is suggested by the fact that
the masses of the two types of objects independently scale with the
luminosity or mass of the host galaxy or bulge, with a similar factor
of proportionality \citep{wehner06, ferrarese06, rossa06}.

A nuclear star cluster and intermediate-mass black hole are known to
coexist in at least one object, NGC~4395, a late-type (Sd) galaxy
which hosts a central cluster coincident with a low-luminosity Seyfert
1 nucleus powered by accretion onto a black hole with mass of $\sim 4
\times 10^5$ M$_\odot$ \citep{filippenko03, peterson05}.  Formation of such a
collapsed object is clearly not inevitable in such environments; M33,
a galaxy of similar morphology, hosts a nuclear star cluster, but no
evidence of a significant black hole ($M_\bullet < 1500$ M$_\odot$,
Gebhardt et al. 2001; $M_\bullet < 3000$ M$_\odot$, Merritt et
al. 2001).  Additional investigation is needed if we are to understand
whether a real connection exists between nuclear star clusters and the
formation of intermediate-mass black holes.  Late-type galaxies such
as NGC~4395 and M33 hold additional special interest as {\em
bulgeless} systems that nonetheless harbor central mass concentrations,
and in the case of NGC~4395, a black hole.  Evidence for the coexistence
of nuclear clusters with black holes as traced by AGNs in galaxies
in general has recently been summarized by \citet{seth08}.

In this paper we report on observations that bear on this matter for
the nuclear star cluster in another bulgeless galaxy, NGC~1042
(morphological type SAB(rs)cd, de Vaucouleurs et al. 1991).  The
presence of a central cluster in this galaxy was first revealed in the
{\em Hubble Space Telescope} imaging survey described by
\citet{boeker02}; an $I$-band image along with a surface brightness
profile of the inner galaxy is given in that reference.
Spectroscopic measurements of the cluster stellar population and mass
have been presented by Walcher et al. (2005, 2006).  Here we focus on
emission-line properties of the galaxy nucleus, and evidence they
provide for accretion power indicative of a central black hole.

\section{Observations}

For our analysis we make use of several observations of the NGC 1042
nucleus.  The spectrum of this source was obtained with the TWIN
spectrograph at the Calar Alto 3.5m telescope on 2005 August 04.
Two grating settings were used to cover a total
bandpass of 3440 -- 7570 \AA\ with a 1\arcsec\ slit, yielding a
spectral resolution of $\sim 3300$.  The seeing during the
observations was $\sim$0\farcs 8 and the slit was oriented at PA=25$^\circ$
which was close to the parallactic angle.  The spectra were calibrated using
standard methods.  Conditions at the time were nonphotometric, and we
determined the scalefactor necessary to place our spectra on an
absolute photometric scale through comparison with a spectrum of
NGC~1042 obtained by the {\em Sloan Digital Sky Survey}
\citep[SDSS;][]{york00}.  For this purpose we used a 3\arcsec-wide
extraction from the TWIN spectrum, which is similar in dimension to
the 3\arcsec-diameter SDSS fiber aperture.  The two spectra are very
similar in continuum and emission-line properties.  The continuum and
emission-line fluxes are spatially strongly peaked; we
consequently did not attempt to correct for the different aperture
areas, but note that the resulting flux scale could be uncertain by as
much as a factor of 2.  (We note that this uncertainty does not affect 
our subsequent conclusions that are based on line ratios.)

The TWIN spectrum of the NGC~1042 nucleus as measured through a
1\arcsec $\times$ 1\arcsec synthetic aperture is shown in Figure 1.
The spectrum shows emission lines superposed on a stellar continuum,
with signal-to-noise ratio decreasing at the blue end.  The aperture
is substantially larger than the central star cluster, which has an
effective radius of $\sim 0$\farcs 02 \citep{boeker04}, and thus the
spectrum in Figure 1 contains significant circumnuclear starlight 
\citep[81\% of the total light in the I bandpass;][]{walcher06}
and nebular emission.

\begin{figure}
\figurenum{1}
\plotone{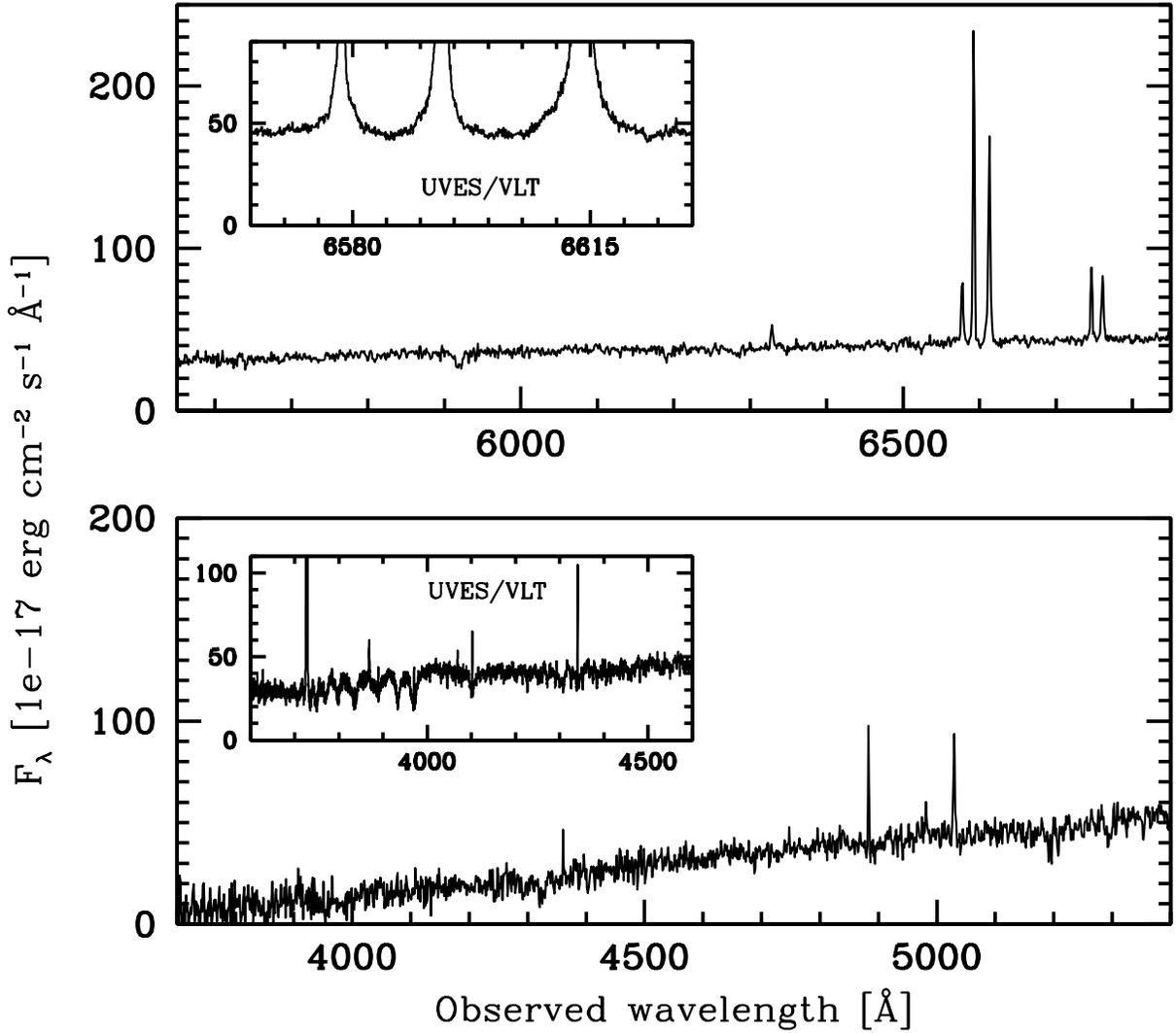}

\caption{Spectra of the nucleus of NGC~1042 obtained with a 1\arcsec
$\times$ 1\arcsec aperture.  The TWIN spectrum is shown with full
wavelength coverage, while portions of the high-resolution UVES
spectra are shown as insets.  The red UVES spectrum is expanded and
truncated vertically to show the profiles of the [\ion{N}{2}]
$\lambda$6548, H$\alpha$ $\lambda$6563, and [\ion{N}{2}]
$\lambda$6583 lines.  Note that the H$\alpha$ profile shows a larger
peak flux but narrower base than the [\ion{N}{2}] lines.}
\end{figure}

We also obtained high resolution spectra for the NGC~1042 nucleus
using UVES at the Very Large Telescope, for purposes of resolving
absorption features and enabling accurate estimation of the stellar
velocity dispersion and population constraints.  The UVES observations
were obtained with a 1\arcsec-wide slit, and span two wavelength
intervals, 3570 -- 4830 \AA\ and 6120 -- 7980 \AA, with resolution of
$\sim35,000$.  The seeing for these observations was $\sim$0\farcs 8,
and we used an extraction width of 1\arcsec to measure the spectrum of
the nucleus.  The instrument slit was maintained at the parallactic
angle throughout, thereby sweeping through Position Angles $123 -
141^\circ$.  Portions of the resulting spectra are shown as insets in
Figure 1.  Further details and analysis of the stellar continuum as
measured in the UVES data are presented by Walcher et al.  (2005,
2006).  As discussed by \citet{walcher06}, the continuum is 
dominated by an old stellar population.

\section{Analysis}

Within the optical bandpass, nebular emission may signal the presence
of an active galactic nucleus through emission-line strengths, ratios,
and/or profiles.  With the data described in \S 2 we are able to
investigate each of these aspects of the NGC~1042 nucleus.

On first inspection, the emission-line ratios for this source are
ambiguous in their interpretation.  We measured the flux of
prominent emission features in the TWIN spectrum; in so doing we
did not attempt to subtract the starlight but corrected the Balmer
emission-line fluxes using the equivalent widths (EWs) of underlying
absorption as determined from the population fit to the 
high-resolution UVES spectra \citep{walcher06}.  The 
resulting line fluxes are listed in Table 1, and line
ratios commonly used for nebular classification are plotted in Figure
2. In each case the ratios for the nucleus fall close to the 
AGN/starburst boundary.  Also shown in Table 1 and Figure 2 are measurements
obtained through adjacent, off-nucleus apertures, spanning $1.5 - 3$\arcsec
on both sides of the nucleus.  The off-nucleus spectrum displays
emission consistent with high-metallicity \ion{H}{2} regions 
photoionized by hot stars.  The nuclear H$\alpha$/H$\beta$ ratio 
is consistent with the Case B prediction ($\sim 2.86$) and hence 
negligible reddening within the measurement uncertainties.

\begin{deluxetable}{lcc}
\tablewidth{0pt}
\tablenum{1}
\tablecaption{NGC~1042 Emission Line Fluxes}
\tablehead{
\colhead{Line} & \multicolumn{2}{c}{$100 \times F/F({\rm H}\beta)$}
\\ & \colhead{Nucleus} & \colhead{Off-Nucleus}}
\startdata
{[\ion{O}{2}]} $\lambda$3726 & $111\pm 22$ & ... \\
{[\ion{O}{2}]} $\lambda$3729 & $ 89\pm 22$ & ... \\
{[\ion{Ne}{3}]} $\lambda$3869 & $35\pm 3$  & ... \\
{\ion{He}{1}} + H8 $\lambda$3889     & $ 12\pm  2$   & ... \\
H$\epsilon$ $\lambda$3970  & $ 14\pm  2$   & ... \\
{[\ion{S}{2}]} $\lambda$4069 & $ 19\pm 5$  & ... \\
H$\delta$ $\lambda$4102    & $ 21\pm  3$   & ... \\
H$\gamma$ $\lambda$4340    & $ 41\pm 12$ & $ 20\pm 12$ \\
H$\beta$ $\lambda$4861     & $100\pm 14$ & $100\pm 14$ \\
{[\ion{O}{3}]} $\lambda$4959 & $ 13\pm 3$ & $ 0\pm  3$ \\
{[\ion{O}{3}]} $\lambda$5007 & $ 84\pm 9$ & $26\pm  8$ \\
{[\ion{O}{1}]} $\lambda$6300 & $ 26\pm 1$ & $0.0\pm 0.6$ \\
{[\ion{N}{2}]} $\lambda$6548 & $ 68\pm 2$ & $39\pm  2$ \\
H$\alpha$ $\lambda$6563    & $332\pm 16$ & $318\pm 15$ \\
{[\ion{N}{2}]} $\lambda$6583 & $234\pm 8$ & $118\pm 6$ \\
{[\ion{S}{2}]} $\lambda$6716 & $ 68\pm 3$ & $44\pm  2$ \\
{[\ion{S}{2}]} $\lambda$6731 & $ 78\pm 3$ & $30\pm  3$ \\
\enddata
\tablewidth{9truecm}
\vspace{-0.8truecm}
\tablecomments{$F({\rm H}\beta) = 1.9\times 10^{-15}$ erg s$^{-1}$ cm$^{-2}$
for the nucleus and $1.8\times 10^{-15}$ erg s$^{-1}$ cm$^{-2}$ for the 
combined off-nucleus apertures.  The absolute flux scale is uncertain by 
up to a factor of 2.  H$\gamma$ and longer wavelength transitions were
measured from the TWIN spectra, while lines shortward of 4800 \AA\ were 
measured from UVES.  The UVES and TWIN measurements were placed on a consistent
flux scale by multiplying the UVES values to produce agreement in the
H$\gamma$ line flux.}
\end{deluxetable}

\begin{figure}
\figurenum{2}
\plotone{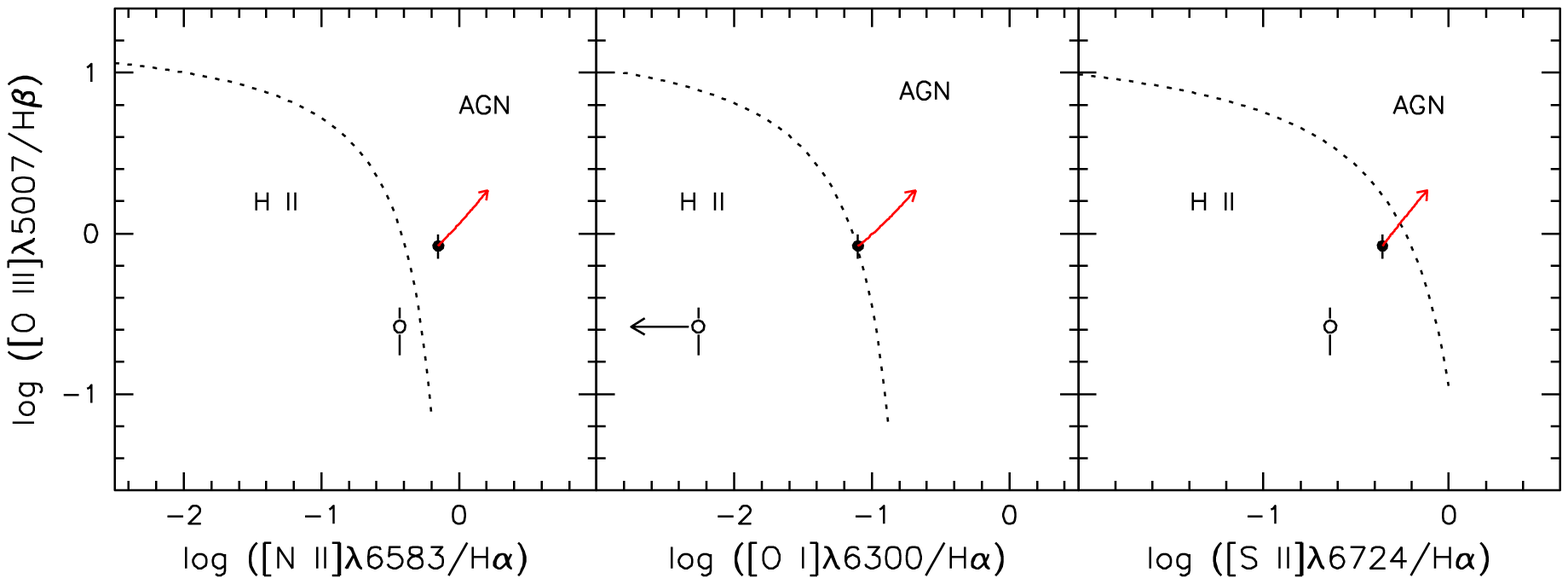}
\caption
{Emission-line flux ratio diagrams for discrimination between
stellar-photoionized and accretion-powered sources.  [\ion{S}{2}]
$\lambda$6724 represents the sum of $\lambda\lambda$6716, 6731.  The
dotted lines separating \ion{H}{2} regions and AGNs are taken from
\citet{kewley01} and \citet{kauffmann03}.  The filled points
represent measurements of the nucleus from the TWIN spectrum through
a 1\arcsec $\times$ 1\arcsec aperture. Open points represent
off-nucleus averages for regions along the slit spanning $1.5 -
3$\arcsec on both sides of the nucleus.  Error bars are not shown 
when smaller than the data points.  The arrows associated with the
filled points illustrate how the
ratios for the nucleus change after removal of contaminating
\ion{H}{2} region emission (see text for details).}
\end{figure}

If we suppose that the nucleus itself is powered by stars, we can
perform a consistency test by comparing the ionizing photon production
rate necessitated to explain the Balmer emission with the ionizing
photon production rate predicted for the observed stars.
\citet{walcher06} obtained detailed multi-component stellar population
fits to the blue UVES continuum spectra, demonstrating that this
source was dominated by old ($\geq$ Gyr) stars with additional younger
components, some with ages $\sim 10^7$ yrs.  The \citet{bruzual03}
models used by Walcher et al. provide quantitative predictions for the
ionizing luminosity emitted by the stars producing the observed blue
spectra.  For the UVES spectra no absolute flux calibration exists,
hence the ionizing photon flux derived from the blue setting and the
H$\alpha$ flux from the red setting are not directly
comparable. However, consistent predictions for the H$\alpha$ emission
line and the adjacent continuum can be obtained from the fits to the
blue spectra, so that the predicted H$\alpha$ emission equivalent
width (EW) can be directly compared with the observed value.  The
predicted EW(H$\alpha$) assuming complete absorption of the ionizing
photons and Case B recombination is 16.2 \AA , which is in excellent
agreement with the measured value of $16.5 \pm 0.3$ \AA .  (The latter
value includes a correction for underlying stellar absorption with EW
= $4.5 \pm 0.3$\AA .)  An accretion source is thus not required to explain
the nebular recombination flux if the stellar ionizing radiation is
absorbed efficiently, and if this radiation truly arises from stars;
in composite optical spectra it can, however, be very difficult to
distinguish hot stellar components from power-law emission powered by
accretion.  We thus cannot exclude the possibility that the blue
continuum component identified in the stellar populations fitting is
in fact a weak AGN continuum that extends beyond 13.6 eV, thereby
contributing to the nebular ionization.

The UVES spectra provide detailed information on emission line
profiles and the results are surprising.  The upper inset in Figure 1
shows the H$\alpha$ $\lambda$6563, [\ion{N}{2}] $\lambda\lambda$6548,
6583 lines, which are the strongest transitions in the red UVES
spectrum.  The [\ion{N}{2}] lines have full width at half maximum
(FWHM) velocities of $\sim 80$ km s$^{-1}$, which is consistent with
the stellar velocity dispersion in the same aperture of $32 \pm 5$ km
s$^{-1}$ \citep{walcher05} for a Gaussian distribution (FWHM =
2.35$\sigma$).  The lines clearly deviate from Gaussian profiles in
that they show prominent wings, most visible in [\ion{N}{2}]
$\lambda$6583, extending to approximately $\pm 300$ km s$^{-1}$ from
line center.  While these large gas velocities are suggestive of AGN
activity, Figure 1 presents a conundrum in that the H$\alpha$ line is
obviously {\em narrower} than the [\ion{N}{2}] lines, in contrast with
Seyferts and LINERs where the lines have comparable widths, or broader
widths for the Balmer lines.  This difference can be seen more clearly
in Figure 3, which shows an overlay of the two profiles.  The
H$\alpha$ line has FWHM $\approx 50$ km s$^{-1}$ which is actually
less than would be predicted from the stellar $\sigma$ in the Gaussian
case.  Underlying stellar absorption does not significantly distort
the H$\alpha$ profile; Figure 3 also shows the result of subtracting
the continuum obtained from the stellar population fit described by
\cite{walcher06}, and the emission profile is substantially unaltered.

\begin{figure}
\figurenum{3}
\epsscale{.50}
\plotone{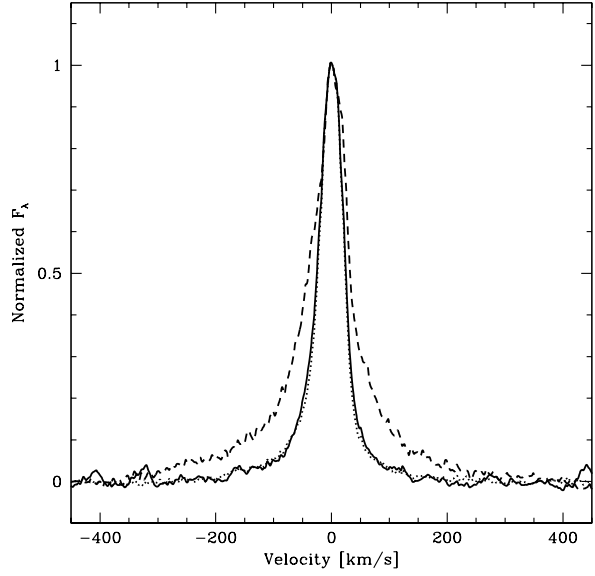}
\caption{Normalized emission profiles measured in the center of NGC~1042
for [\ion{N}{2}] $\lambda$6583 (dashed) and H$\alpha$ (dotted) after 
subtraction of a constant continuum level.  The residual H$\alpha$ 
profile after subtraction of a model stellar continuum is shown by the 
solid line.}
\end{figure}

Insight into the unusual line profiles can be gained by examining the
spatial variation of emission properties along the UVES slit.  The
emission-line flux as a function of position is plotted in Figure 4,
which shows measurements obtained from {1\arcsec}-wide extractions
stepped along the slit.  The figure shows that the H$\alpha$ peak flux
is offset by $\sim$ 0\farcs 5 (i.e. $\sim 40$ pc) from the location of the
peak flux in the forbidden lines and in the stellar continuum.
This offset is likewise evident in a published SAURON H$\beta$ map
\citep{ganda06}.
Notably, the location of maximum forbidden line flux is also the
location of maximum velocity width for {\em all} of the lines,
including H$\alpha$ (Figure 5); at the same site, the
[\ion{O}{1}]/H$\alpha$, [\ion{N}{2}]/H$\alpha$, and
[\ion{S}{2}]/H$\alpha$ ratios achieve maxima. Figure 4 shows that the
[\ion{S}{2}] $\lambda$6716/[\ion{S}{2}] $\lambda$6731 flux ratio also
exhibits spatial variations, with a minimum value indicative of a
maximum plasma density at the site of peak forbidden-line emission.
It is additionally noteworthy that emission in [\ion{O}{1}], which tends to
trace nonstellar ionization processes, is the most highly
concentrated, consistent with a compact source centered at the
location of peak line and continuum flux.  Inspection of the rotation
curve measured from the emission lines (Figure 6) indicates that 
the position of peak continuum and forbidden-line flux is consistent
with the kinematic center for this system, although an offset of $\sim 12$
km s$^{-1}$ is seen between H$\alpha$ and [\ion{O}{1}] in the central 
aperture.

\begin{figure}
\figurenum{4}
\epsscale{.50}
\plotone{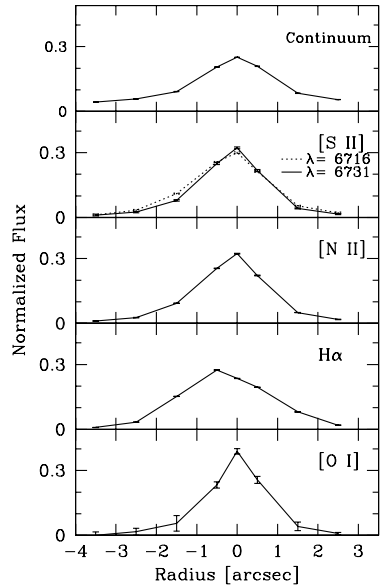}
\caption{Emission-line flux as a function of position along the slit as
measured in the UVES spectra with 1\arcsec-wide extractions.  The spatial
profiles are normalized such that the area under each curve is equal, with
the exception of [\ion{S}{2}] $\lambda$6716, which is scaled to the same
normalization as [\ion{S}{2}] $\lambda$6731.}
\end{figure}

\begin{figure}
\figurenum{5}
\epsscale{0.50}
\plotone{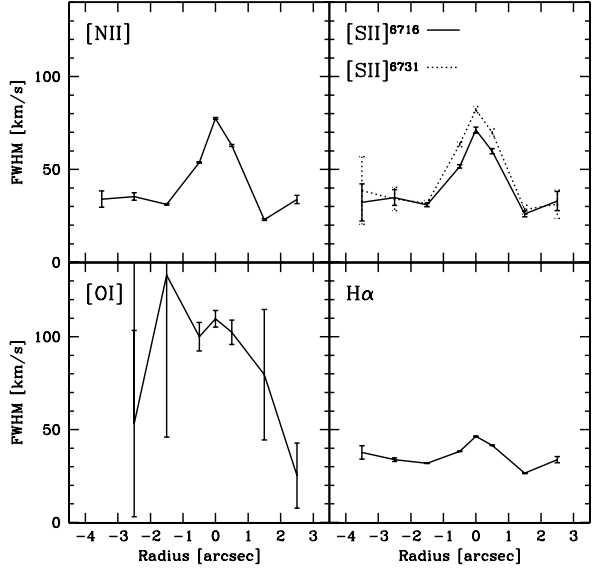}
\caption{Emission-line FWHM as a function of position along the slit
as measured in the UVES spectra with 1\arcsec-wide extractions.}
\end{figure}

\begin{figure}
\figurenum{6}
\epsscale{0.50}
\plotone{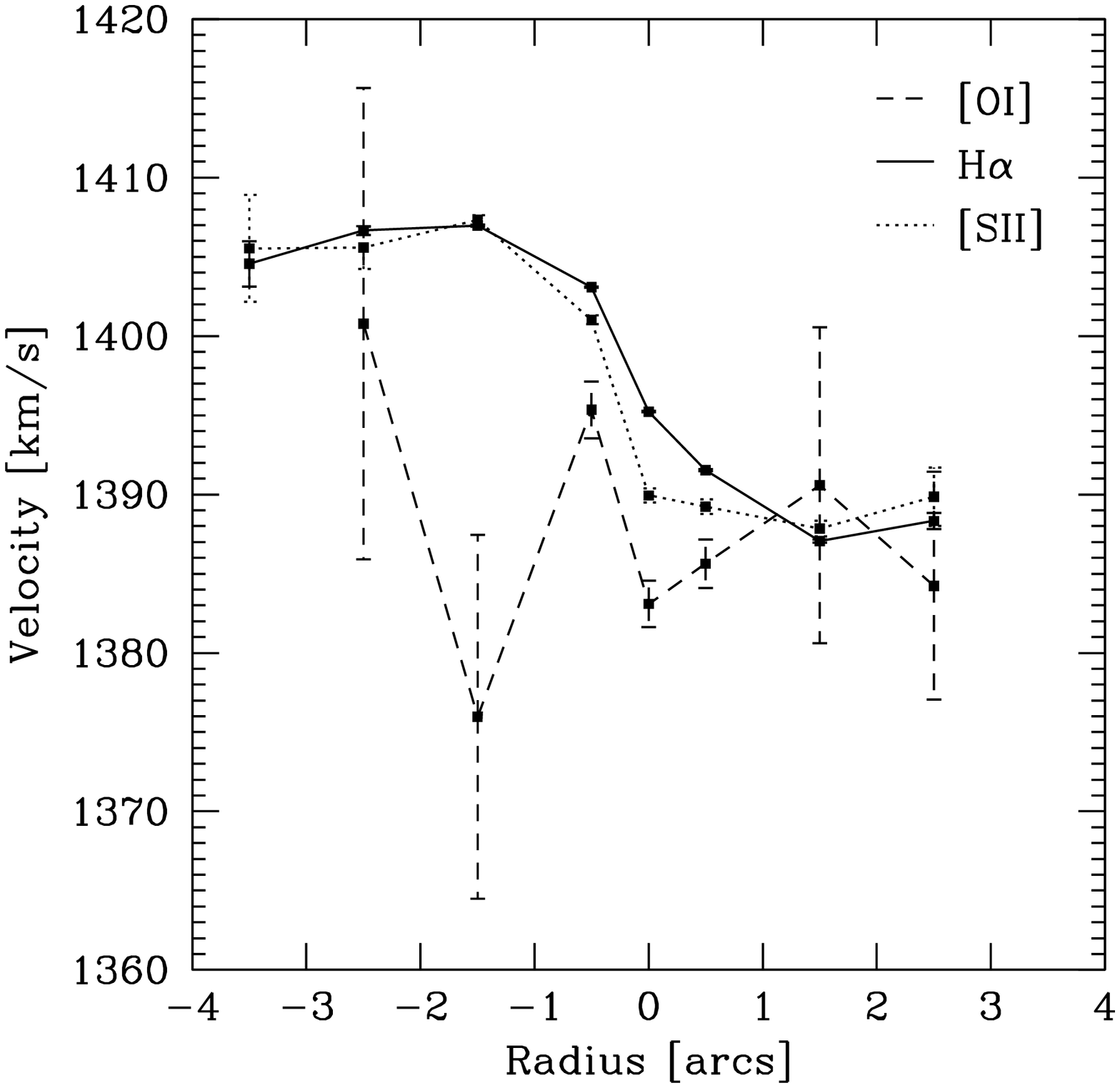}
\caption{Emission-line radial velocity as a function of position 
along the slit as measured in the UVES spectra with 1\arcsec-wide 
extractions.  Velocities are determined from Gaussian fits to the
line profiles.}
\end{figure}

The spatially resolved UVES results can be understood collectively if
an accretion source is located at the site of maximum line width,
forbidden-line flux, and continuum emission -- i.e., coincident with
the nuclear star cluster, as well as the kinematic center -- and an
offset \ion{H}{2} region is responsible for the peak H$\alpha$ flux.
The line profiles seen in the central extraction shown in Figure 1 are
thus naturally explained as the blended emission from a weak AGN and
an \ion{H}{2} region, with the latter component dominating the
H$\alpha$ line.  The composite nature of the nucleus additionally
provides an explanation for why line ratios plotted in Figure 2 fall
near the AGN/starburst demarcation.  We obtained an approximate
decomposition of the two contributions by assuming that the
[\ion{N}{2}] $\lambda$6583 profile was dominated by the accretion
source.  We first subtracted the underlying stellar continuum from the
H$\alpha$ + [\ion{N}{2}] wavelength region for the nucleus using the
multi-age composite stellar fit described by \citet{walcher06}.  We
then scaled the [\ion{N}{2}] profile to match the H$\alpha$ wings, and
subtracted the scaled profile to determine approximately what fraction
of H$\alpha$ remains that can be attributed to the \ion{H}{2} region,
i.e., 61\%.  We can use this fraction to correct the emission ratios
shown in Figure 2, and for this purpose use the emission spectrum from
the off-nucleus aperture as a template for the \ion{H}{2} component.
(This estimate assumes that H$\alpha$ and H$\beta$ have the same
relative proportions of AGN and \ion{H}{2} region emission in the
central aperture.)  The resulting shifts in position for the ratios
for the central region are shown in Figure 2; in each case the central
source moves more definitively into the locus of AGNs.

While the detailed decomposition described here is only approximate
and cannot be considered unique, it is clear that removal of H$\alpha$
and other emission associated with the source adjacent to the nucleus
(Figure 4) would offset the line ratios to values in stronger accord
with a central accretion source.  Similar examples of mixed AGN and
\ion{H}{2} emission revealed through line profile differences are
described by, e.g., \citet{veron81} and \citet{shields90}.  In NGC~1042
the underlying AGN is a type 2 object (no detected broad H$\alpha$)
with emission-line ratios consistent with classification as a LINER.

High energy emission would be expected from the AGN, but will require
deep observations in order to be detectable.  An archival {\sl ROSAT} HRI
image exists for NGC~1042, with an exposure of 22 ksec in 1995 July.
No source is evident at the position of the nucleus, and taking into
consideration the Poisson nature of a possible source and local
background yields a 90\%-confidence upper limit of 7.4 counts for the
AGN.  Assuming a distance to NGC~1042 of 18.2 Mpc, absorption by the
Galactic column density of $3.1 \times 10^{20}$ cm$^{-2}$
\citep{dickey90}, and an intrinsic power-law source with photon index
$\Gamma = 2$, this limit translates into $L(0.1 - 2.4\, {\rm keV}) < 7.9
\times 10^{38}$ erg s$^{-1}$. Extrapolating the same assumed continuum
to harder energies would imply $L(2-10\, {\rm keV}) < 4.0 \times
10^{38}$ erg s$^{-1}$.  It is useful to combine the latter value with
the AGN H$\alpha$ luminosity, for comparison with other systems.
After removing the estimated contamination from the adjacent
\ion{H}{2} region (see \S 3) and correcting for Galactic extinction of
$A_V = 0.095$ mag \citep{schlegel98}, the H$\alpha$ luminosity for the
NGC~1042 nucleus is $8.0 \times 10^{37}$ erg s$^{-1}$.  The upper
limit to the x-ray luminosity is easily consistent with measurements
for other weak AGNs with comparable $L({\rm H}\alpha)$ \citep{ho01}.
The lack of a detection with {\sl ROSAT} is thus consistent with the
low luminosity of the AGN indicated by its optical emission.

\section{Discussion}

From analysis of the NGC 1042 optical spectra we conclude that this source
presents strong evidence for an accreting black hole coincident with
a compact nuclear star cluster.  We can draw several other inferences
from this source which are of interest for comparison with other
galaxies.

While we have no means of measuring the black hole mass, we can place
a lower limit on its value if the source's luminosity is less than the
Eddington luminosity.  We can obtain an order-of-magnitude estimate of
the radiative output of the AGN using measurements of the H$\alpha$
line.  If this low luminosity source has a
spectral energy distribution similar to typical LINERs, then we expect
the bolometric luminosity to be $L_{bol} \approx 100 L({\rm H}\alpha)$
\citep{ho99, ho01}.  The H$\alpha$ measurement thus implies $L_{bol}
\approx 8\times 10^{39}$ erg s$^{-1}$ for NGC 1042, and hence
$M_\bullet \gtrsim 60$ M$_\odot$ if the AGN accretes at a
sub-Eddington rate.

An upper bound on $M_\bullet$ can also be obtained from measurements
of the nuclear star cluster based on estimates of the mass-to-light
ratio $M/L$ for this source.  If $M_\bullet$ is significant in
comparison with the stellar mass in the central cluster, $M/L$ based
on dynamical estimates of total mass will be increased by a
corresponding amount.  \citet{walcher05} obtained a dynamical
estimate of $M/L$ for the nucleus of NGC~1042, and \citet{walcher06}
reported an independent estimate of $M/L$ using
multi-component stellar population fits to the optical spectra.  The
latter method will not include the contribution from any dark mass,
but Walcher et al.  find that $M/L$ obtained from the stellar
population analysis is formally somewhat {\em larger} than the
dynamical estimate.  We can conclude from this that $M_\bullet$ must
be modest in comparison with the cluster stellar mass, $3 \times 10^6$
M$_\odot$ \citep{walcher05}, and can thus adopt this value as an
upper limit for the black hole.

We can estimate black hole mass from the stellar $\sigma$ using the
$M_\bullet - \sigma$ relation \citep{gebhardt00, ferrarese00},
although its applicability in this case is uncertain.  The
$M_\bullet - \sigma$ relation is based on measurements of $\sigma$ for
host galaxy bulges, while in NGC~1042 we only have a measure of
$\sigma$ for the central star cluster.  In this context it is
noteworthy that measurements of the globular cluster G1 in M31 are
consistent with extrapolation of the relation to the scale of massive
globular star clusters \citep{gebhardt05}.  For NGC~1042 the
$M_\bullet - \sigma$ relation as derived by \citet{tremaine02}
would predict $M_\bullet \approx 4 \times 10^5$ M$_\odot$.  This value
is consistent with the bounds obtained above, and would imply a rather
low value of the Eddington ratio $L/L_{Edd} \approx 2 \times 10^{-4}$, 
in accord with the LINER classification for the observed AGN
\citep{ho04}.

Based on these arguments, we can state that an intermediate-mass black
hole with $60$ M$_\odot \lesssim M_\bullet \lesssim 3\times 10^6$ M$_\odot$
resides in
NGC~1042's nucleus, coincident with the central star cluster.  This
discovery provides added incentive to consider star clusters as an
essential element in spawning collapsed objects that grow into the
central engines of luminous AGNs.  NGC~1042 thus may present some of
the attributes today of early galaxies that have evolved to host
supermassive black holes.  Regardless of the detailed evolutionary 
pathway, the formation of massive compact star clusters and black holes
requires substantial dissipation and collapse of matter to the central
parsecs of a galaxy.  NGC~1042, along with NGC~4395, provide
demonstrations that the two types of massive central objects can
readily coexist, a point that should be considered when linking
these structures in a common framework \citep[e.g.,][]{ferrarese06, wehner06}.

The relationship between $M_\bullet$ and other galaxy attributes in
objects like NGC~1042 remains ill-determined.  It is natural to ask
whether the central black hole in such galaxies is linked to the
nuclear cluster in some manner analogous to the black hole-bulge
correlations found in earlier Hubble types.  However, the situation
remains confused in that the other known examples, NGC~4395 and M33 
(\S 1), fall off the extrapolated $M_\bullet - \sigma$ relation with
$M_\bullet$ values that are too high and too low, respectively, for
the $\sigma$ as measured in the central region\footnote{$M_\bullet$
for NGC~4395 obtained from reverberation mapping \citep{peterson05} is
larger than expected from the $M_\bullet - \sigma$ relation, but
$M_\bullet$ estimates obtained by other means, as described by 
\citet{filippenko03}, result in better agreement.}.  Several authors have
suggested that $M_\bullet$ is linked most fundamentally to galaxy
halo mass, rather than bulge mass \citep[e.g.,][]{ferrarese03, baes03},
but this idea remains controversial \citep{ho07}.
Bulgeless galaxies like NGC~1042 offer an interesting class of objects 
for probing these ideas \citep[see also][]{satyapal07, satyapal08}.
One complication is the finding by \citet{boeker03} that 
apparently bulgeless galaxies often have modest central light excesses
above a simple extrapolation of an exponential disk.  In this
regard it is worth noting that NGC~1042 has one of the largest such
excesses among the galaxies analyzed by B{\"o}ker et al.  It is tempting
to regard this finding as further evidence linking central black holes
and bulge-like structures, but a more systematic survey is needed to 
draw definite conclusions.

\section{Conclusions}

In this study we have presented evidence from optical spectroscopy
that a low-excitation AGN resides in the nucleus of the late-type galaxy
NGC~1042.  Existing constraints suggest that the accretion activity is
powered by an intermediate-mass black hole.  This result is noteworthy
in that a compact star cluster also resides in the nucleus;
consequently this source may implicate an important role for star
clusters in generating ``seed'' black holes in galaxy centers.
NGC~1042 is also remarkable in hosting a central black hole while
lacking a stellar bulge.  Understanding the implications for black
hole formation in relation to galaxy evolution will require study of
statistical samples of similar galaxies, where simultaneous
measurements of bulge, central cluster, and black hole properties are
possible.

\acknowledgments

JS and CJW thank the Max-Planck-Institut f{\"u}r Astronomie for
support for visits enabling much of this work to be completed.  CJW is
supported by the MAGPOP Marie Curie EU Research and Training Network.
This research is based in part on observations collected at the
European Southern Observatory, Chile [ESO Programme 68.B-0076(A)], and
at the Centro Astron{\'o}mico Hispano Alem{\'a}n at Calar Alto,
operated jointly by the Max-Planck-Institut f{\"u}r Astronomie and the
Instituto de Astrof{\'\i}sica de Andaluc{\'\i}a.  We thank Reynier
Peletier and Anil Seth for informative conversations, and Nicolas
Cardiel for making his software package {\sl Reduceme} available along
with expert advice on how to use it.  We are grateful to Alexej
Kniazev for access to service observing time at Calar Alto which
enabled this project to be completed.

\end{document}